\documentclass[conference]{IEEEtran}
\IEEEoverridecommandlockouts
\usepackage{amsmath,amssymb,amsfonts}
\usepackage{algorithmic}
\usepackage{graphicx}
\usepackage{array}
\usepackage{textcomp}
\usepackage[hidelinks]{hyperref}
\usepackage{multirow} 
\usepackage{placeins}
\usepackage{float}
\usepackage{gensymb}
\usepackage{xcolor}
\usepackage[x11names]{xcolor}
\usepackage{subcaption}
\usepackage{cite}
\usepackage[font=small,labelfont=bf]{caption}
\def\BibTeX{{\rm B\kern-.05em{\sc i\kern-.025em b}\kern-.08em
    T\kern-.1667em\lower.7ex\hbox{E}\kern-.125emX}}
\begin{document}

\title{Linear State Estimation in Presence of Bounded Uncertainties: A Comparative Analysis
}

\author{\IEEEauthorblockN{Ayan Das, \textit{Student Member}, IEEE}
\IEEEauthorblockA{\textit{Department of Electrical Engineering} \\
\textit{Indian Institute of Technology Bombay}\\
Mumbai, India \\
210070016@iitb.ac.in}
\and
\IEEEauthorblockN{Anushka Sharma, \textit{Student Member}, IEEE, \\ Anamitra Pal, \textit{Senior Member}, IEEE}
\IEEEauthorblockA{\textit{School of Electrical, Computer, and Energy Engineering} \\
\textit{Arizona State University},
Tempe, USA \\
ashar561@asu.edu, 
anamitra.pal@asu.edu}
}

\maketitle

\begin{abstract}
A variety of algorithms have 
been proposed to address the power system state estimation problem in the presence of uncertainties in the data. However, less emphasis has been given to handling perturbations
in the model. In the context of linear state estimation (LSE), which is the focus of this paper, perturbations in the model come from variations in the line parameters. 
Since the actual values of the line parameters can be different from the values stored in a power utility's database, we investigate three approaches in this paper to estimate the states 
in the presence of \textit{bounded} uncertainties in the data and the model. The first approach is based on interval arithmetic, the second is based on convex optimization, and the third 
is based on generalized linear-fractional programming.
The three algorithms are applied to multiple IEEE test systems and compared in terms of their speed and accuracy. The results indicate that the first two algorithms are extremely fast and give expected results,
while the third suffers from scalability issues and is unsuitable for LSE.
\end{abstract}


\section{Introduction}
Power system state estimation is an active area of research, owing to the need for accurate knowledge of bus voltages in real-time.
As such, considerable focus has been devoted to phasor measurement unit (PMU)-based state estimation, which can provide an estimate of the states (bus voltage phasors) at sub-second timescales \cite{10495872}.
The simplest time-synchronized state estimation formulation involving PMUs is the linear state estimation (LSE) problem that is usually solved using least squares-based approaches.
However, the LSE problem becomes 
more 
complicated (a) when the measurements
are corrupted by bad data/noise having arbitrary attributes, (b) in the presence of topology variations and unknown transformer tap-ratios, and/or (c) if the measurement data is insufficient 
\cite{10026224,9930859,9535418,9275590,7374723,cheng2021adaptive}.
For these scenarios, more advanced
state estimation algorithms have been developed, depending on the location and type of the complication. 

To minimize the impact of bad data/outliers, robust state estimation formulations based on least-absolute value and least-modulus, have been proposed for PMU-only state estimators \cite{gol2014robust,sun2019optimum,jabr2024complex}.
To account for topology variations, either a topology identifier is run before/in combination with the PMU-only state estimator \cite{kim2013on,azimian2022state,zhou2025adaptive}, or a topology-change resilient state estimator is created \cite{moshtagh2023time,moshtagh2025topology}.
Similarly, the impact of unknown tap-ratios is minimized by incorporating them into the state estimation process \cite{fernandes2017application,cheng2024survey}.
Lastly, to tackle scarcity of measurement data for performing PMU-only state estimation, pseudo-measurements and machine learning have been employed \cite{glavic2013reconstructing,azimian2021time,azimian2024analytical}.

It is clear from this literature review 
that considerable attention has been given to addressing uncertainties in the data.
In contrast, uncertainties in the model have not received as such attention (apart from considering topology variations).
For the PMU-only LSE problem, a major source of \textit{bounded model uncertainty} is the change in the line parameter values caused by environmental conditions (e.g., temperature), operating conditions (heavy/light loading), and aging \cite{varghese2023transmission}.
Note that the line parameters can change by as much as $\pm 30\%$ from their nominal values stored in a power utilities' database \cite{kusic2004measurement}, implying that their impacts on PMU-only LSE can be considerable.
In \cite{static}, an extended weighted least square approach was developed for conventional (non-PMU) state estimation to
estimate 
bus voltage phasors as well as parameter errors accurately. However, the computational complexity of this approach 
raised concerns regarding its use in
PMU-only state estimation which is done at sub-second timescales.
Lastly, 
the IEEE/IEC Standard \cite{IEEE-IEC-STD2018} mandates that PMU data can have a maximum
total vector error (TVE) of 1\% under nominal conditions.
This means that a bound also exists
on the uncertainty in the PMU data used for doing LSE.

This paper advances the state-of-the-art by comparing
the speed and accuracy of three 
algorithms that can do
PMU-only LSE in the presence of bounded uncertainties in line parameters (model) and PMU measurements (data).
The first approach is based on \textit{interval arithmetic}, the second on \textit{convex optimization}, and the third 
on \textit{generalized linear-fractional programming} (GLFP).
A special emphasis is given to the computational cost of the three approaches 
by evaluating them explicitly for standard IEEE test systems.
Possible applications and drawbacks of the three approaches 
are also discussed.


\section{Overview of Three Algorithms that Account for Uncertainties in Linear Regression Models}
\label{Review}

The LSE problem investigated in this paper has the following relationship:
\begin{equation}
\boldsymbol{y}=\boldsymbol{P}(\boldsymbol{p})\boldsymbol{x}-\boldsymbol{\delta y}
\label{eq1}
\end{equation}
where $\boldsymbol{p}$ represents the line parameters, and $\boldsymbol{y}$ consists of the phasor measurements (voltage and current phasors obtained from PMUs placed at optimal locations in the grid \cite{pal2017pmu,pal2014pmu}) converted into rectangular coordinates with the real and the imaginary parts stacked one below the other.
The measurement matrix, $\boldsymbol{P}$, which relates the measurements with the states, is composed of the line parameters, and is hence a function of $\boldsymbol{p}$. 
The states of the system, which are to be estimated, are denoted by $\boldsymbol{x}$, while
$\boldsymbol{\delta y}$ is the bounded PMU measurement noise.
The line parameters possess bounded uncertainties, and therefore elements of $\boldsymbol{P}$ are erroneous.
The following subsections provide a summary of the three mathematical algorithms 
that are used in this paper to solve the LSE problem described by \eqref{eq1}.


\subsection{State Estimation Bounds Using Interval Arithmetic}
\label{Interval Arithmetic State Estimation}

The LSE problem obeying \eqref{eq1} 
can be 
expressed as a weighted least squares optimization problem
whose goal is to minimize the objective function, $\psi(\boldsymbol{x})$, given by:
\begin{equation}
    \psi(\boldsymbol{x}) = 
    (\boldsymbol{y} - \boldsymbol{P(p)}\boldsymbol{x})^T \boldsymbol{W}^{-1} (\boldsymbol{y} - \boldsymbol{P(p)x})
    \label{eq2}
\end{equation}

In \eqref{eq2}, $\boldsymbol{W}$ is a diagonal matrix constructed from the inverse of the variances in the PMU measurements. The value of $\boldsymbol{x}$ that minimizes $\psi(\boldsymbol{x})$ can be obtained by solving the following system of linear equations \cite{interval14}:
\begin{equation}
    \begin{bmatrix}
        \boldsymbol{P(p)} & \boldsymbol{-1} \\
        \boldsymbol{0} & \boldsymbol{P^T(p)W^{-1}}
    \end{bmatrix}
    \begin{bmatrix}
        \boldsymbol{\hat{x}} \\
        \boldsymbol{y}_d
    \end{bmatrix}
    =
    \begin{bmatrix}
        \boldsymbol{y} \\
        \boldsymbol{0}
    \end{bmatrix}
    \label{eq3}
\end{equation}
where $\boldsymbol{y}_d$ is a dummy vector, and $\boldsymbol{1}$ and $\boldsymbol{0}$ are identity and zero matrices, respectively.
Now, the uncertainty in the $k^{th}$ network parameter, $p_k$, can 
be modeled as
\cite{interval}: 
\begin{equation}
p_k=p_{k_{nom}}+\Delta p_k\epsilon_k
\label{eq4}
\end{equation}
where $\Delta p_k$ represents the maximum deviation in the nominal line parameter value $p_{k_{nom}}$, and $\epsilon_k \in [-1,1]$ is the uncertainty term.
The measurement matrix, $\boldsymbol{P(p)}$, considering network
parameter uncertainty can then be expressed as:
\begin{equation}
\boldsymbol{P(p)} = \boldsymbol{P}_{0m} + \sum_{k=1}^{2n_p} \Delta p_k \boldsymbol{P}_k \epsilon_k
\label{eq5}
\end{equation}
where $n_p$ is the number of parameters under consideration, and $\boldsymbol{P}_{0m}$ is the measurement matrix constructed from the nominal values of the line parameters. Now, \eqref{eq3} can be re-written as:
\begin{equation}
\left( \boldsymbol{A}_{0m} + \sum_{k=1}^{2n_p} \boldsymbol{A}_k \Delta p_k \epsilon_k \right)
\begin{bmatrix}
\hat{\boldsymbol{x}} \\
\boldsymbol{y}_d
\end{bmatrix}
=
\begin{bmatrix}
\boldsymbol{y} \\
\boldsymbol{0}
\end{bmatrix}
\label{eq6}
\end{equation}
where 
\begin{equation}
\left.
\begin{aligned}
\boldsymbol{A}_{0m} = 
\begin{bmatrix}
\boldsymbol{P}_{0m} & -\boldsymbol{1} \\
\boldsymbol{0} & \boldsymbol{P}_{0m}^T \boldsymbol{W}^{-1}
\end{bmatrix}\!; \, \, \,
\boldsymbol{A}_k = 
\begin{bmatrix}
\boldsymbol{P}_k & \boldsymbol{0} \\
\boldsymbol{0} & \boldsymbol{P}_k^T \boldsymbol{W}^{-1}
\end{bmatrix}
\end{aligned}
\right.
\label{eq7}
\end{equation}

The limits on
the estimated values of $\boldsymbol{x}$ under parameter uncertainties
can be found by solving the following optimization problems for $j=1,...,2B$, where $B$ is number of buses:
\begin{equation}
\begin{aligned}
\underline{x}_j^\star &= \min \; \boldsymbol{e}_j^T \boldsymbol{x} \\
\text{subject to} \quad & 
\left\{
\begin{array}{l}
\text{Eq. (6)}, \\
|\epsilon_k| \leq 1; \quad (k = 1, \ldots, 2n_p)
\end{array}
\right.
\end{aligned}
\label{eq8}
\end{equation}
\begin{equation}
\begin{aligned}
\overline{x}_j^\star &= \max \; \boldsymbol{e}_j^T \boldsymbol{x} \\
\text{subject to} \quad & 
\left\{
\begin{array}{l}
\text{Eq. (6)}, \\
|\epsilon_k| \leq 1; \quad (k = 1, \ldots, 2n_p)
\end{array}
\right.
\end{aligned}
\label{eq9}
\end{equation}
where $\boldsymbol{e}_j$ is a vector of appropriate dimensions with the $j^{th}$ element equal to 1 and the rest of the elements equal to 0.

In accordance with \cite{interval}, interval arithmetic is 
used to solve \eqref{eq8} and \eqref{eq9} to find $\underline{\boldsymbol{x}}^*$ and $\overline{\boldsymbol{x}}^*$. Using interval numbers and assuming $\boldsymbol{A}_{0m}^{-1}$ exists, we let $\boldsymbol{C}_k = \boldsymbol{A}_{0m}^{-1} \boldsymbol{A}_k \Delta p_k$ and $\boldsymbol{f}_{0m} = \boldsymbol{A}_{0m}^{-1} \boldsymbol{b}_z$, where $\boldsymbol{b}_z^T = \begin{bmatrix} \boldsymbol{y}^T & \boldsymbol{0} \end{bmatrix}$. Then, the outer solution is given by $[\boldsymbol{f}] = \boldsymbol{f}_{0m} + \boldsymbol{u} [-1, 1]$ with $\boldsymbol{u}>0$, and \eqref{eq6} becomes:
\begin{equation}
    \boldsymbol{u}[-1, 1] + \sum_{k=1}^{2n_p} \boldsymbol{C}_k \boldsymbol{f}_{0m}[\epsilon_k] + \boldsymbol{C}_k \boldsymbol{u}[-1, 1][\epsilon_k] = \boldsymbol{0}
\label{eq10}
\end{equation}
where $[-1,1][\epsilon_k] \in [-1,1]$. 
To determine $u$, we define $\boldsymbol{u}^{(0)}=0$, $j=0$,
and $[\boldsymbol{w}] = -\sum_{k=1}^{2n_p} \boldsymbol{C}_k \boldsymbol{f}_{0m}[\epsilon_k]$, and perform the following iteration until it converges:
\begin{equation}
    \boldsymbol{u}^{(j+1)}[-1, 1] = [\boldsymbol{w}] - \sum_{k=1}^{2n_p} \boldsymbol{C}_k \boldsymbol{u}^{(j)}[-1, 1]
\label{eq11}
\end{equation}

Finally, 
the bounded uncertainty in the $i^{th}$ measurement can be accounted for in the following way:
\begin{equation}
    \delta y_i = \Delta y_i \epsilon_i
\label{eq12}
\end{equation}
where $\Delta y_i$ is the maximum deviation in $\delta y_i$, and $\epsilon_i \in [-1,1]$ is the uncertainty term.
The outer solutions are now estimated by modifying $[\boldsymbol{w}]$ in \eqref{eq11} as follows:
\begin{equation}
    [\boldsymbol{w}] = \sum_{i=1}^{n} \boldsymbol{A}_{0m}^{-1} \boldsymbol{e}_i \Delta y_i \epsilon_i [-1,1] - \sum_{k=1}^{2n_p} \boldsymbol{C}_k \boldsymbol{f}_{0m}[\epsilon_k]
\label{eq13}
\end{equation}
where $n$ is the number of elements in $\boldsymbol{y}$, and $\boldsymbol{e}_i$ is a vector of appropriate dimensions with the $i^{th}$ element equal to 1 and the rest of the elements equal to 0.

Note that the interval arithmetic-based solution does not provide the actual state estimate but rather the range inside which the estimates lie.
The next two approaches provide a single value of the state estimate
in the presence of bounded uncertainties in network parameters and PMU measurements. 



\subsection{State Estimation Using Convex Optimization}
\label{ConvexProg}

Since the bounds on the line parameters are finite, the uncertainty in $\boldsymbol{P(p)}$, denoted by $\boldsymbol{\delta P}$, is also finite/bounded.
Let the induced-2 norm of the bound on $\boldsymbol{\delta P}$
be $\chi_P$.
Similarly, let the bound on $\boldsymbol{\delta y}$ be $\chi_y$. Then,
the states can be estimated from \eqref{eq1} by solving the following $\mathrm{min}$-$\mathrm{max}$ problem:

\vspace{-1em}

{\small
\begin{equation}\min_{\boldsymbol{\hat{x}}} \max\Big\{\| (\boldsymbol{P}+\boldsymbol{\delta P}) \boldsymbol{\hat{x}}-(\boldsymbol{y}+\boldsymbol{\delta y})\|_2: \|\boldsymbol{\delta P}\|_2\leq\chi_P, \|\boldsymbol{\delta y}\|_2\leq\chi_y\Big\}
\label{eq14}
\end{equation}}

The $\mathrm{min}$-$\mathrm{max}$ problem in \eqref{eq14} can be converted into an equivalent convex optimization
problem as shown below:
\begin{equation}
\min_{\boldsymbol{\hat{x}}} (\boldsymbol{\|P\hat{x}-y}\|_2+\chi_P\|\boldsymbol{\hat{x}}\|_2+\chi_y)
\label{eq15}
\end{equation}

This convex minimization problem is solved by defining \eqref{eq15} as a cost function to
obtain the following result \cite{chandra}:
\begin{equation}\boldsymbol{\hat{x}}=\begin{pmatrix}\boldsymbol{P^TP}+\theta \boldsymbol{I}\end{pmatrix}^{-1}\boldsymbol{P^Ty}
\label{eq16}
\end{equation}
where $\theta$, 
given by \eqref{eq17}, is computed mathematically by solving a secular equation (see Section 3 of \cite{chandra} for more details):
\begin{equation}
\theta=\frac{\chi_P \boldsymbol{\|P\hat{x}-y}\|_2}{\|\boldsymbol{\hat{x}}\|_2}
\label{eq17}
\end{equation}



\subsection{State and Uncertainty Bound Estimation using Generalized Linear-Fractional Programming (GLFP)}
For the quantities following the linear 
relationship shown in \eqref{eq1}, 
let 
the uncertainty in the elements of $\boldsymbol{P}$ be
bounded by $\boldsymbol{\xi}$\footnote{Note that $\boldsymbol{\xi}$ 
is the limit of the additive uncertainty in the elements of $\boldsymbol{P}$, while $\chi_P$, mentioned in Section \ref{ConvexProg}, was
the induced-2 norm of $\boldsymbol{\delta P}$.
}.  
In accordance with \cite{hladik}, for estimating both $\boldsymbol{\xi}$ and $\boldsymbol{x}$, the following GLFP problem can be formulated:
\begin{equation}
\boldsymbol{v}_s=\min_{\boldsymbol{x}\in\mathbb{R}^{2B}}\left\{\max_{\substack{i\in\{1,...,n\}\\k\in\{0,1\}}}\left.\frac{(-1)^{1-k}\boldsymbol{l}_i^T\boldsymbol{x}+(-1)^k\boldsymbol{y}_i}{\boldsymbol{\zeta^T}\boldsymbol{D}_s\boldsymbol{x}+1}\right|\boldsymbol{D}_s\boldsymbol{x}\ge\boldsymbol{0}\right\}
\label{eq18}
\end{equation}
where $B$ is the number of buses, $\boldsymbol{s}$ is a sign vector $\{\pm1\}^{2B}$ having 
length 
equal to twice the number of buses, and $\boldsymbol{D}_s$ is $\mathrm{diag}(\boldsymbol{s})$. The column vector $\boldsymbol{\zeta}$ defined as shown below:
\begin{equation}
\zeta_i=\left\{\begin{array}{ll}1&\quad \text{if} \;i \in \boldsymbol{\Lambda}\\0&\quad \text{if} \;i \notin \boldsymbol{\Lambda}\end{array}\right.
\label{eq19}
\end{equation}

In the above equations, column vector
$\boldsymbol{l}_i$ is the $i^{th}$ row of $\boldsymbol{P}$, $y_i$ is the $i^{th}$ element of $\boldsymbol{y}$, and $\boldsymbol{\Lambda}$ is a known matrix that indicates which regressors are known precisely and which have uncertainties.
Following this, we can write:
\begin{equation}\widehat{\boldsymbol{\xi}}=\min_{\boldsymbol{s}\in\{\pm1\}^{2B}}\boldsymbol{v}_s
\label{eq20}
\end{equation}
and if $s^*$ is $\mathrm{argmin}$ of \eqref{eq20}, then the optimal state estimate
is the corresponding 
$\boldsymbol{x}_s^*$.

Note that unlike the convex optimization-based solution described in Section \ref{ConvexProg}, the solution obtained by the GLFP approach estimates both the states as well as the bounds on the uncertainties in the elements of $\boldsymbol{P}$.
However, because of the exponential complexity 
(in terms of the number of buses, $B$)
of the GLFP approach, its computational cost/burden is much higher than the other two approaches.

\section{Implementation Methodology}
\label{Implementation}
The three algorithms described in Section \ref{Review}
were implemented using MATPOWER \cite{matpower_soft}.
Data for the simulations was generated by
solving the power flows after modifying the line parameters appropriately.
Using this data and the equations provided in the previous section, the LSE problem given by \eqref{eq1} was solved.  
The IEEE 14, 30, 57, and 118-bus systems were used for the interval arithmetic and convex optimization-based approaches.
Because of its very high computational burden, the GLFP approach could only be performed for the IEEE 5-bus system (its results for this system were compared with the interval arithmetic approach).
For the 14, 30, 57, and 118-bus systems, PMUs were placed at those locations which ensured complete observability of the respective systems \cite{pal2017pmu,pal2014pmu}. For the 5-bus system, PMUs were placed on all five buses.


\subsection{Data Generation}
\label{PMU Measurements}

The original line parameters were modified using the \texttt{randn} function in MATLAB, while ensuring
that the deviations
lie within the $\pm$30\% range. 
Next, the actual line parameter
values 
were replaced with the 
perturbed values to create new case files in MATPOWER. 
The power flow was solved
using the new case files to produce the voltage and current phasors. 
A Gaussian noise bounded by
1\% TVE 
was then added to the phasors to create the PMU measurements.
To mimic the real-world scenario in which the line parameter information in the power utilities' database is outdated, the $\boldsymbol{P}$ matrix was constructed using the original line parameter values. This made the $\boldsymbol{P}$ matrix erroneous w.r.t. the PMU data that was obtained using the modified system.  

\subsection{Obtaining Bounds on State Estimates Using Interval Arithmetic}
With the input data generated using the steps outlined in Section \ref{PMU Measurements}, the standard deviation of the difference between the true and the noisy measurements 
was calculated, and the covariance matrix, $\boldsymbol{W}$, constructed from it. For implementing interval arithmetic, the CORA library \cite{cora} was used.

\subsection{Estimating States 
Using Convex Optimization}
The noisy PMU measurements obtained from the power flow analysis of the system with perturbed line parameters were 
used to create
$\boldsymbol{y}$, while the $\boldsymbol{P}$ matrix contained the original line parameter values.
The value of $\chi_P$
was set to be the maximum of the absolute value of the difference between the perturbed and the unperturbed line parameters.
To solve the secular equation, \texttt{fsolve}, available in MATLAB's Optimization Toolbox \cite{Optimization_Toolbox}, was utilized. 

\subsection{Estimating States and Bounds Using GLFP}
This approach could only be implemented on the 5-bus system. The necessary matrices for this test system and approach were generated using the methodology described in Section \ref{PMU Measurements}.
For modeling and solving \eqref{eq18}, the GUROBI Optimization Package \cite{gurobi} was called from within MATLAB. 

\section{Results and Discussion}
\label{Results}
\begin{figure*}[t]
    \centering
    \includegraphics[width=1 \textwidth, height=2.5in]{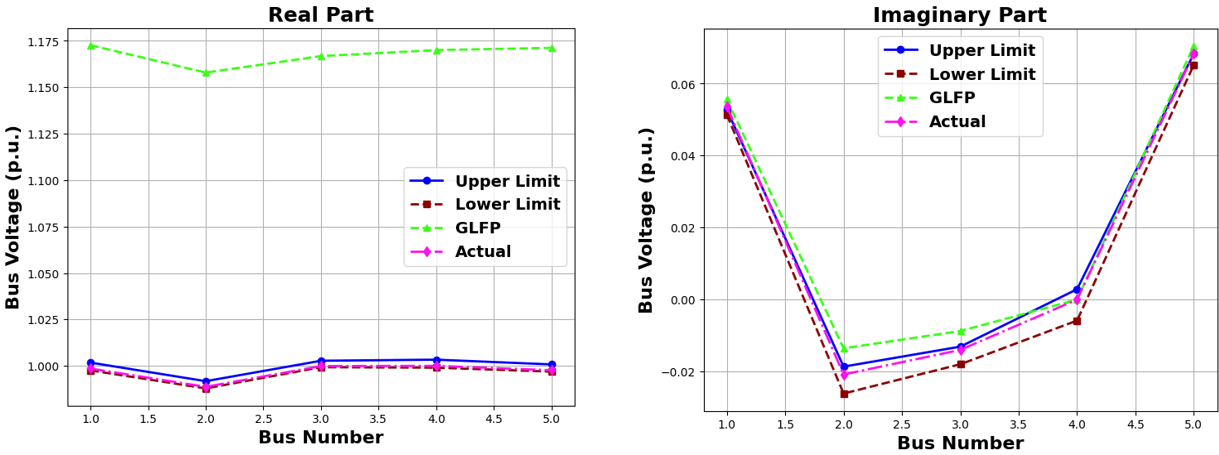}
    \vspace{-1.5em}
    \caption{Estimated voltages for 5-bus system. The voltages estimated using GLFP approach lie outside the bounds, while the actual voltages lie within the bounds obtained using interval arithmetic approach.}
    \label{fig:f5}
\end{figure*}

\begin{figure*}[t]
    \vspace{-0.5em}
    \centering
    \includegraphics[width=1\textwidth, height=2.5in]{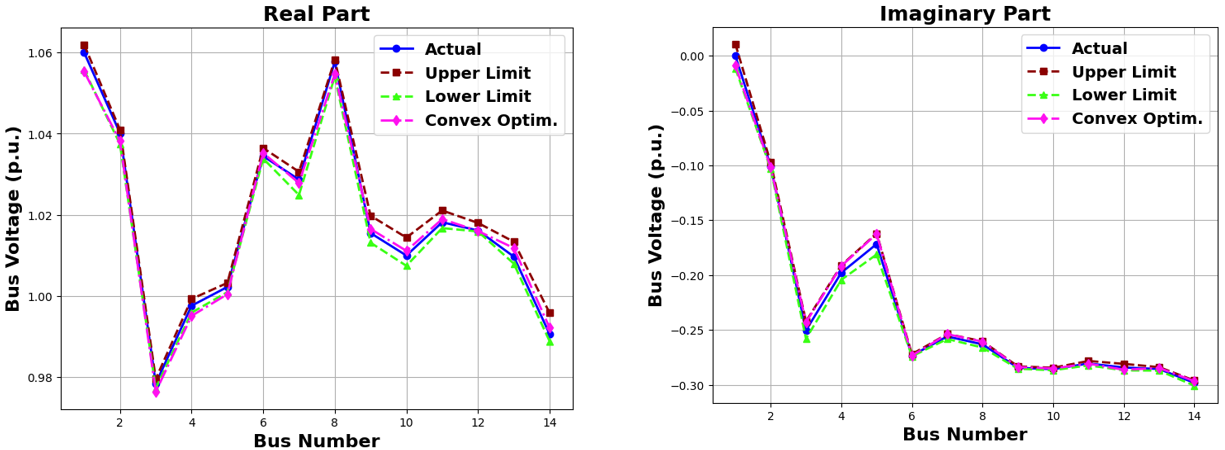}
    \vspace{-1.5em}
    \caption{Estimated voltages for 14-bus system. 
    For the real part as well as the imaginary parts, the actual voltages and the voltages estimated using convex optimization approach, lie within the bounds obtained using interval arithmetic approach.
    }
    \label{fig:f14}
    \vspace{-1em}
\end{figure*}

\begin{figure*}[t]
    \centering
    \includegraphics[width=\textwidth, height=2.5in]{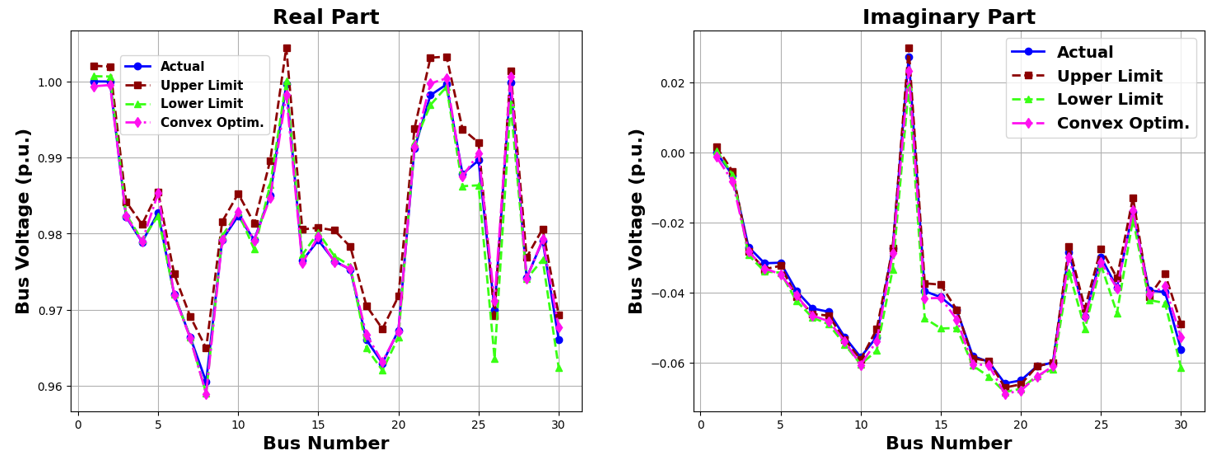}
    \vspace{-1.5em}
    \caption{Estimated voltages for 30-bus system. 
    For real part, all actual voltages and most of the voltages estimated using convex optimization approach lie within the bounds obtained using interval arithmetic approach.
    For imaginary part, most of the actual voltages and all of the voltages estimated using convex optimization approach lie within the bounds obtained using interval arithmetic approach.
    }
    \label{fig:f30}
\end{figure*}

\begin{figure*}[t]
    \vspace{-0.5em}
    \centering
    \includegraphics[width=\textwidth, height=2.5in]{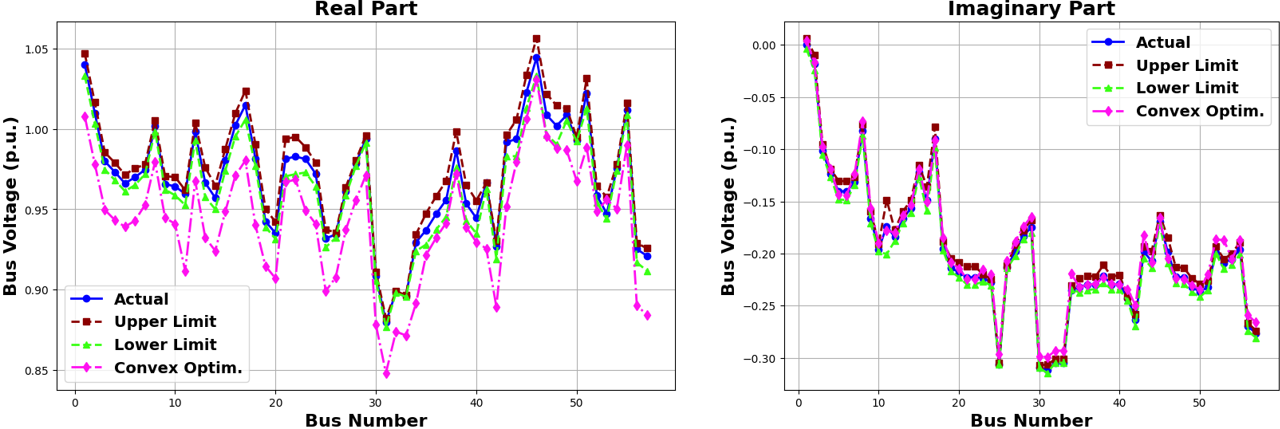}
    \vspace{-1.5em}
    \caption{Estimated voltages for 57-bus system. For real part, actual voltages lie within the bounds obtained using interval arithmetic approach, while voltages estimated using convex optimization approach lie mostly below the lower bound. 
    For imaginary part,
    all actual voltages and most of the voltages estimated using convex optimization approach lie within the bounds obtained using interval arithmetic approach.
    }
    \label{fig:f57}
    \vspace{-1.5em}
\end{figure*}

\begin{figure*}[t]
    \centering
    \includegraphics[width=\textwidth, height=2.5in]{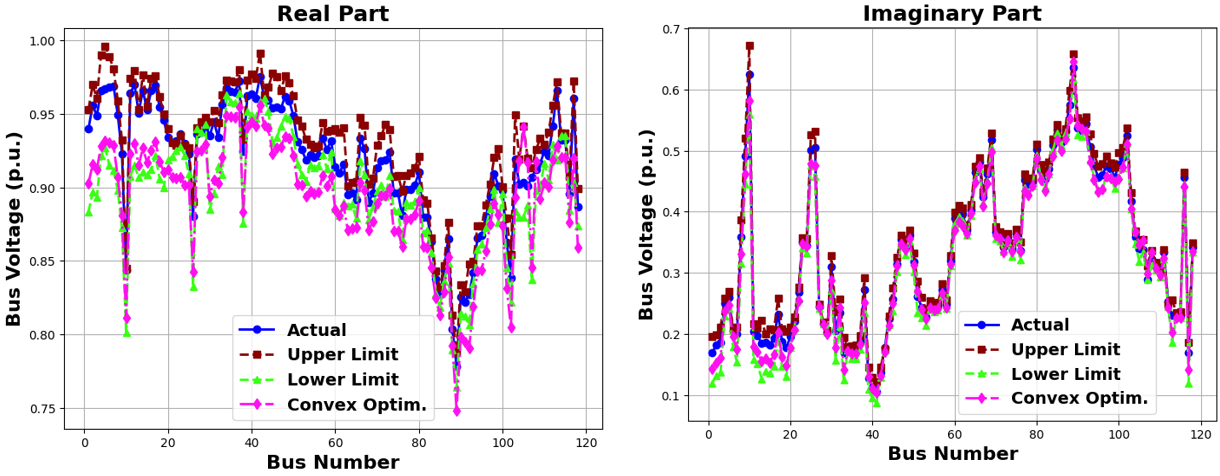}
    \vspace{-1.5em}
    \caption{Estimated voltages for 118-bus system.
    For real part, actual voltages lie within the bounds obtained using interval arithmetic approach, while the voltages estimated using convex optimization approach lie mostly below the lower bound. 
    For imaginary part, 
    all actual voltages and most of the voltages estimated using convex optimization approach lie within the bounds obtained using interval arithmetic approach.
    }
    \label{fig:f118}
    \vspace{-1em}
\end{figure*}

The results 
are presented in  Figs. \ref{fig:f5}-\ref{fig:f118}, which depict the estimates of the real and imaginary parts of the bus voltages for the different test systems.
For the 5-bus system shown in Fig. \ref{fig:f5}, it is evident that although the actual voltages lie within the bounds obtained using the interval arithmetic approach, those obtained using the GLFP approach did not. The reason for this could be the sensitivity of the GLFP approach to the matrix elements and/or the perturbations.
For the voltage estimates obtained using the convex optimization approach for the 14, 30, 57, and 118-bus systems (see Figs. \ref{fig:f14}-\ref{fig:f118}), the following inferences are drawn: (i) the imaginary parts either lay within the bounds or very close to the lower bound obtained by the interval arithmetic approach; (ii) the real parts often lay below the lower bound, particularly for the larger systems. The actual voltages almost always lay within the bounds obtained using


\noindent the interval arithmetic approach, with the bounds being relatively narrow especially for the larger systems.  
It can be concluded from this analysis that 
if a single estimate value is required, the convex optimization approach should be used, while if a narrow bound for the state estimates suffices, then the interval arithmetic approach should be preferred.




The execution times for the convex optimization and interval arithmetic approaches are summarized in Table \ref{tab:results}. It can be observed from the table that the interval arithmetic approach is almost an order of magnitude faster than the convex optimization approach. 
The root mean squared error (RMSE) results for the convex optimization algorithm is also provided in Table \ref{tab:results}. 
The entries of the table correspond to the run that created Figs. \ref{fig:f5}-\ref{fig:f118}. 
For the 5-bus system, the RMSE of the voltage estimated using the GLFP algorithm was 0.121 p.u., while the duration of program execution was 120.28 seconds, making the GLFP algorithm unsuitable for high-speed PMU-only state estimation. Note that all the simulations were performed on a personal computer having 16 GB of RAM and an AMD Ryzen 7 processor without the utilization of any GPU.

\begin{table}[ht]
\caption{Summary of results obtained using the convex optimization and interval arithmetic-based approaches}
\vspace{-0.5em}
\centering
\begin{tabular}{|c|cc|c|}
\hline
\multirow{2}{*}{\shortstack{\textbf{Test} \\ \textbf{System}}} & \multicolumn{2}{c|}{\textbf{Runtime (seconds)}} & \multirow{2}{*}{\shortstack{\textbf{RMSE for} \\ \textbf{Convex Opt. (p.u.)}}} \\
\cline{2-3}
 & \textbf{Convex Opt.} & \textbf{Interval Arith.} & \\ \hline
14  & 0.444 & 0.0398 & 0.00348 \\ \hline
30  & 0.482 & 0.0416 & 0.00155 \\ \hline
57  & 0.555 & 0.038 & 0.02014  \\ \hline
118 & 0.529 & 0.039 & 0.02287  \\ \hline
\end{tabular}
\label{tab:results}
\vspace{-1.5em}
\end{table}

\section{Conclusion}
\label{Conclusion}
Due to outdated line parameter information 
being stored in a power utility's database,
there is a genuine need for a method
that can estimate
bus voltages accurately using the old parameter values, that too within a reasonable timeframe. 
This paper addresses this need by comparing the performance of three mathematical algorithms for doing
LSE in the presence of bounded perturbations in PMU measurements and
line parameter values. 
The interval arithmetic-based algorithm
provided bounds of the voltages in less than \textcolor{black}{0.05} seconds for all four test systems
and is, therefore, extremely fast. 
The convex optimization-based algorithm 
provided solutions within 0.6 seconds
for systems as large as 118 buses, with good RMSE for the smaller systems. 
Despite giving an estimate of the state as well as quantifying the uncertainty, the approach based on GLFP
has the drawback of having an extremely long execution time.
Additionally, this approach was 
sensitive to the values of the matrix elements.
Although the  computational burden of the GLFP algorithm could be reduced through parallelization, its high sensitivity to the elements of the matrix made it unsuitable for PMU-only LSE.

In summary, the interval arithmetic-based approach and the convex optimization-based
approach are suitable
for LSE purposes. Both of these approaches can provide voltage information quickly with faster processors, 
given that the simulations conducted in this paper were on a computer with relatively low processing power. 
The future scope of this work involves comparison with recently proposed machine learning approaches for transmission system state estimation (e.g., \cite{10495872,moshtagh2025topology}) and validation-via-hardware implementation.



\bibliographystyle{IEEEtran}
\bibliography{references.bib}

\end{document}